\documentstyle[12pt]{article}
\begin{document}
\title{BFT Method for Mixed Constrained Systems \\
and Chern-Simons Theory}
\author{M. Monemzadeh$^{a}$\thanks{monemzadeh@sepahan.iut.ac.ir}
\\A. Shirzad$^{a,b}$\thanks{shirzad@ipm.ir}\\  \\
  $^a$~{\it Department of  Physics, Isfahan University of Technology (IUT)}\\
{\it Isfahan,  Iran,} \\
 $^b$~{\it Institute for Studies in Theoretical Physics and Mathematics (IPM)}\\
{\it P. O. Box: 19395-5531, Tehran, Iran.}}
\date{}
\maketitle

\begin{abstract}
We show that the BFT embedding method is problematic for mixed
systems (systems possessing both first and second class
constraints). The Chern-Simons theory as an example is worked out
in detail. We give two methods to solve the problem leading to two
different types of finite order BFT embedding for Chern-Simons
theory.
\end{abstract}
\section{Introduction}
Canonical quantization of constrained systems is fully established
in the framework of Dirac theory \cite{Dirac}. As is well-known,
in the case of second class systems one should convert Dirac
brackets to quantum commutators; while for first class systems one
constructs the quantum space of states as some representation of
all quantized operators (i.e. phase space coordinates) and then
imposes the conditions $\Phi_a |\mbox{phys}> =0$, where $\Phi_a$
are first class constraints and $|\mbox{phys}>$ means physical
states.

Working with first class systems seems to be appealing for some
reasons; firstly, because the symmetries and covariance of the
classical theory are manifestly demonstrated; secondly, since
converting Dirac brackets to quantum commutators sometimes implies
factor ordering problem and quantization of these models is not
formal; thirdly, because inverting the matrix of Poisson brackets
of constraints, which is necessary for writing the Dirac brackets,
is not generally an easy task; and finally the most important
reason is that the construction of a BRST charge is possible only
 for first class systems
\cite{BRST,Henn}. Therefore, there are some efforts to convert a
second class system to a first class one
\cite{BatFrad,BatFrad2,BatTyu}. The method, recognized as the BFT
method, is based on extending the phase space to include a set of
new variables and then writing the constraints, as well as the
physical quantities, as power series in terms of these added
variables.

 However, as we will
explain in the following, the traditional BFT method is formulated
only for pure second class systems \cite{BanBanGh2}, while in the
general case both first and second class constraints may emerge in
the same model. An important example of this case, i.e. {\it mixed
constrained systems}, is the Chern-Simons theory (abelian and
non-abelian). After a brief review in the next section of the
finite order BFT method, as proposed in \cite{monshi1} for a pure
second class system, we will show in section 3 that in fact it is
not possible to embed the second class constraints in a larger
space separately. That is, when one tries to convert the second
class constraints into first class ones via embedding, the algebra
of the original first class constraints may change; in other
words, they will not necessarily remain first class.

We will investigate the origin of this violence and search for
conditions that can guarantee the embedding of second class
constraints without violating the involuting algebra of first
class ones. We show that the non-abelian Chern-Simons theory is a
special example which exhibits this violence. In section 4 we
propose two distinct methods that help us to solve the problem.
The first method concerns redefining the constraints so that their
algebra fulfill the required condition. In the next method we
suggest that at first stage one may convert the first class
constraints into second class ones by means of adding some
auxiliary variables, and then one is able to run the procedure of
the usual BFT method. We will show that this suggestions enables
us to construct  BFT embedding for Chern-Simons theory.
\section{Finite order BFT embedding}
Consider a pure second  class constrained system described by the
Hamiltonian $H_0$ in some phase space with coordinates $(q^i,p_i)$
where $i=1,2,...K$. Assume we are given a set of second class
constraints, $\tau_\alpha^{(0)} \hspace {3mm}\alpha=1,...m$,
satisfying the algebra
 \begin{equation}
 \Delta_{\alpha\beta}=\left\{\tau_\alpha^{(0)}, \tau_\beta^{(0)}\right\}
 \label{a1}
 \end{equation}
where $ \left\{ , \right\}$ means Poisson bracket and
$\Delta_{\alpha\beta}$ is an invertible matrix. To convert this
second class system into a gauge system, i.e. a first class
system, one should extend the phase space by introducing the same
number of auxiliary variables
 as that of second class constraints. We denote these variables by
$\eta_{\alpha}$ and assume that they obey the following algebra;
 \begin{equation}
 \left\{\eta_{\alpha}, \eta_{\beta}\right\}= \omega_{\alpha\beta}
 \label{a2}
 \end{equation}
where $\omega_{\alpha\beta}$ is an antisymmetric invertible matrix
which may be proposed arbitrarily. The first class constraints in
the extended phase space $(q,p)\oplus\eta$ are defined as
 \begin{equation}
 \tau_{\alpha}(q,p,\eta)=\sum_{n=0}^\infty \tau_\alpha^{(n)}
  \hspace {1cm} \alpha=1,2,....,m
 \label{a3}
 \end{equation}
  where $\tau^{(n)}_\alpha$ is of order $n$ with respect to
$\eta_\alpha$'s and
\begin{equation}
 \tau^{(0)}_\alpha = \tau_{\alpha}(q,p,0).
 \label{a7}
 \end{equation}
In the abelian BFT embedding method one demands that these
extended constraints be strongly involuting:
 \begin{equation}
 \left\{\tau_{\alpha},\tau_{\beta}\right\}=0.
 \label{a5}
 \end{equation}
Substituting Eq. (\ref{a3}) into Eq. (\ref{a5}) leads to a set of
recursive relations. Vanishing of the term independent of $\eta$
gives:
\begin{equation}
 \left\{\tau^{(0)} _\alpha , \tau^{(0)} _\beta\right\}+ \left\{\tau^{(1)} _\alpha , \tau^{(1)}
  _\beta\right\}_{(\eta)}=0;
  \label{a8}
  \end{equation}
and vanishing of the term of order $n$ with respect to
$\eta_\alpha$'s for $n\geq1$ gives
  \begin{equation}
  \left\{\tau_{[\alpha}^{(1)},\tau_{\beta ]}^{(n+1)}\right\}_{(\eta)}+B_{\alpha\beta}^{(n)}=0 \hspace{1cm}
  n\geq 1
  \label{a9}
  \end{equation}
where
  \begin{eqnarray}
  B_{\alpha\beta}^{(1)}&\equiv& \left\{\tau_{[\alpha}^{(0)},
  \tau_{\beta]}^{(1)}\right\},
  \label{b10}\\
  B_{\alpha\beta}^{(n)}&\equiv&
  \frac{1}{2}B_{[\alpha\beta]}\equiv \sum_{m=0}^n \left\{\tau_\alpha ^{(n-m)}, \tau_\beta
  ^{(m)}\right\}+\sum_{m=0}^{n-2} \left\{\tau_\alpha^{(n-m)}, \tau_\beta ^{(m+2)}\right\}_{(\eta)}\hspace
  {1cm} n\geq2.
  \label{a10}
  \end{eqnarray}
The suffix $\eta$ in the above equations means that the Poisson
brackets must be evaluated with respect to $\eta$ variables only,
otherwise they are calculated in the basis $(q,p)$. The above
equations are used iteratively to obtain the correction terms
$\tau^{(n)}$. Since $\tau^{(1)}$  is linear with respect to $\eta$
we may write
 \begin{equation}
 \tau_\alpha ^{(1)}={\chi_{\alpha}}^{\beta} (q,p)\eta_\beta.
 \label{a11}
 \end{equation}
Substituting this expression into Eq.(\ref{a8}) and using
Eqs.(\ref{a1}) and (\ref{a2}) we obtain:
 \begin{equation}
 \Delta_{\alpha\beta}+{\chi_{\alpha}}^{\gamma}\omega_{\gamma\lambda}
 {\chi_{\beta}}^{\lambda}=0.
 \label{a12}
 \end{equation}
This equation contains two sets of unknown elements;
${\chi_{\alpha}}^{\beta}$ and $\omega_{\alpha\beta}$. One should
at first assume a suitable anti-symmetric matrix for
$\omega_{\alpha\beta}$ and then solve Eq. (\ref{a12}) to determine
the coefficients ${\chi_{\alpha}}^{\beta}$. Since
$\Delta_{\alpha\beta}$ and $\omega_{\alpha\beta}$ are
anti-symmetric matrices, there exist totally $\frac{m(m-1)}{2}$
independent equations for ${\chi_{\alpha}}^{\beta}$, while the
number of ${\chi_{\alpha}}^{\beta}$'s is $m^2$. Therefore, an
infinite number of solutions for ${\chi_{\alpha}}^{\beta}$ can be
found and we are allowed to chose any solution we wish. Using this
possibility, ${\chi_{\alpha}}^{\beta}$'s can be chosen such that
the process of determining the correction terms $\tau^{(n)}$
terminates at this stage, i.e. $\tau^{(2)}$ vanishes. We will come
to this point later. It can be shown \cite{BanBanGh2,BanBanGh1}
that the general solution of Eq. (\ref{a9}) is given by
 \begin{equation}
 \tau_\alpha ^{(n+1)}=-\frac{1}{n+2} \eta_\mu
 \omega^{\mu\nu}{{\chi^{-1}}_{\nu}}^{\rho}B_{\rho\alpha}^{(n)}; \hspace {1cm}n\geq1
 \label{a13}
 \end{equation}
where $\omega^{\alpha\beta}$ and ${{\chi^{-1}}_{\alpha}}^{\beta}$
are inverse of $\omega_{\alpha\beta}$ and
${\chi_{\alpha}}^{\beta}$ respectively.

To construct  the corresponding Hamiltonian $H(q,p,\eta)$ in the
extended phase space we demand
  \begin{equation}
  H=\sum_{n=0}^\infty\tilde{H}^{(n)}
  \label{a16}
  \end{equation}
  such that
   \begin{equation}\begin{array}{l}
 H(q,p,0)=H_0(q,p)    \\
   \left\{\tau_\alpha,H\right\}=0,
  \label{a15}\end{array}
  \end{equation}
where $H^{(n)}$ is of order $n$ with respect to $\eta_\alpha$'s.
Substituting from Eqs. (\ref{a3}) and (\ref{a16}) in the second
line of Eq. (\ref{a15}) gives:
   \begin{equation}
   \left\{\tau_\alpha^{(1)}, H^{(n+1)}\right\}_{(\eta)}+
   G_\alpha^{(n)}=0;\hspace {1cm}
   n\geq0
   \label{a18}
   \end{equation}
where $G_\alpha^{(n)}$ as the generators of the $H^{(n+1)}$ are
defined as the following
\begin{eqnarray}
G_\alpha ^{(0)} &\equiv& \left\{\tau_\alpha ^{(0)},
H^{(0)}\right\} \label{a19} \\
G_\alpha ^{(1)}&\equiv& \left\{\tau_\alpha ^{(1)},
H^{(0)}\right\}+ \left\{\tau_\alpha ^{(0)},
H^{(1)}\right\}+\left\{\tau_\alpha ^{(2)},
H^{(1)}\right\}_{(\eta)}  \label{a20}\\
G_\alpha ^{(n)}&\equiv& \sum_{m=0}^n  \left\{\tau_\alpha ^{(n-m)},
H^{(m)}\right\}+ \sum_{m=0}^{n-2}   \left\{ \tau_\alpha ^{(n-m)},
H^{(m+2)}\right\}_{(\eta)}+\left\{\tau_\alpha^{(n+1)},
H^{(1)}\right\}_{(\eta)}; \ n\geq2.\ \ \ \   \label{a21}
\end{eqnarray}
It can be shown that the general expression for $H^{(n)}$ is
 \begin{equation}
 H^{(n+1)}=-
 \frac{1}{n+1}\eta_\mu
 \omega^{\mu\nu}{{\chi^{-1}}_{\nu}}^{\lambda}G^{(n)}_\lambda.
 \label{a22}
 \end{equation}

This completes the BFT method of converting a second class system
to a strongly involuting first class one. As can be seen the
correction terms $\tau_\alpha ^{(n)}$ and $H^{(n)}$ are derived
iteratively from Eqs. (\ref{a13}) and (\ref{a22}). Generally,
there is no guarantee that the series terminate at some definite
order. However, the series will terminate if
$B_{\alpha\beta}^{(n)}$ and $G_\alpha^{(n)}$ vanish for a certain
order $n$. If the $\Delta$-matrix in (\ref{a1}) is constant this
goal can be reached simply. In this case it is easily seen that
the choice
\begin{equation} \begin{array}{l}
\omega=-\Delta \\ \chi= \textbf{1} \label{s12}
\end{array} \end{equation}
solves the basic equation (\ref{a12}). With this choice we have
$\tau_{\alpha}^{(1)}=\eta_{\alpha}$ and $B_{\alpha\beta}^{(1)}=0$
(see Eq. \ref{b10}). Then from Eq. (\ref{a10}) all other
$B_{\alpha\beta}^{(n)}$ for $n>1$ vanish. This leads to the
following finite order embedding for the constraints
\begin{equation}
 \tau_\alpha=\tau_{\alpha}+\eta_{\alpha}.
 \label{s13}
 \end{equation}
One can show that in this case the embedding series for
Hamiltonian will also truncate provided that $H^{(0)}$ be a
polynomial function of phase space coordinates \cite{monshi1}.
\section{The problem with mixed systems}
Consider a mixed constrained system which is described by the
Hamiltonian $H^{(0)}(q,p)$. Suppose the system possesses a set of
first class constraints $\phi_i$ as well as the second class ones
$\tau_\alpha^{(0)}$. The problem is to find an embedding in such a
way that the extended Hamiltonian $H$, and the extended
constraints $\tau_\alpha$ and $\tilde{\phi_i}$ have vanishing
Poisson brackets altogether. In other words in addition to Eqs.
(\ref{a5}) and (\ref{a15}) we expect that
\begin{equation}
\left\{H , \tilde{\phi_i}\right\}=0 .
 \label{d1}
\end{equation}
The set of Eqs. (\ref{a5}),(\ref{a15}) and (\ref{d1}) should be
solved simultaneously. It may seem that an embedding for the
second class constraints suffices; i.e. one may consider
$\tilde{\phi_i}$ the same as $\phi_i$ and extend only
$\tau_\alpha^{(0)}$ and $H^{(0)}$ into $\tau_\alpha$ and $H$
respectively. The point is that in general there is no guarantee
that the first class constraints remain still first class. In
other words, the constraints $\phi_i$ may no longer have
vanishing Poisson brackets with the embedded Hamiltonian. To see
this better, suppose in the original theory the secondary first
class constraints, $\phi_{s}$, have been emerged from the
consistency of some primary first class constraints, $\phi_{p}$.
Since in the embedded model some terms should be added to the
Hamiltonian, it is possible that the Poisson brackets
$\{\phi_{p},H\}$ may no more give the same $\phi_{s}$. They may
have been changed to $\tilde{\phi}_{s}$ such that the new set of
constraints $\phi_{p}$ and $\tilde{\phi}_{s}$ are second class.
Therefore, the process of embedding may destroy the gauge
symmetry generated by the set of first class constraints
$\phi_{p}$ and $\phi_{s}$.

Now let us go through the details to see when this may happen. We
know from Eq. (\ref{a16}) that $\tilde{\phi_i}=\phi_i$ will solve
Eq. (\ref{d1}) if
\begin{equation}
\left\{H^{(n)},\phi_i \right\} =0.
 \label{s4}
\end{equation}
 Considering Eq. (\ref{a22}) for a finite order BFT embedding in
 which $\omega_{\alpha\beta}$ and ${\chi_{\beta}}^{\nu}$ are chosen as
 in Eqs. (\ref{s12}), shows that Eq. (\ref{s4}) will be satisfied if
\begin{equation}
 \left\{G_\alpha^{(n)} , \phi_i\right\}=0.
  \label{d3}
  \end{equation}
For $n=0$ we have from Eq. (\ref{a19})
\begin{equation}
\left\{ G_\alpha^{(0)} , \phi_i \right\}=\left\{ \left\{
\tau^{(0)}_\alpha , H_c\right\} , \phi_i \right\}.
 \label{h2}
\end{equation}
In a second class system the Poisson brackets of constraints with
the canonical Hamiltonian vanish weakly except for the constraints
of last level. This may be better understood in chain by chain
approach \cite{LoranShir1}, where the constraints are collected as
chains and within each chain the consistency of every constraint
gives the next one, i.e.
\begin{equation}
 \left\{ \tau^{(0)}_{\alpha-1} , H_c\right\} =\tau^{(0)}_\alpha
 \hspace{1.5cm}\alpha=1,\cdots A.
 \label{s1}
\end{equation}
Since $\tau^{(0)}_{\alpha}$ are second class, at the last level
$\tau^{(0)}_{A}$ should have non-vanishing Poisson bracket at
least with one of the primary constraints. However, nothing can be
said about $\left\{ \tau^{(0)}_A , H_c\right\} $; it may vanish,
may be constant or may be any function of phase space coordinates
which may or may not commute with first class constraints
$\phi_{i}$. Therefore, one way to guarantee Eq. (\ref{d3}) for
$n=0$ is to demand that
\begin{equation}
 \left\{ \tau^{(0)}_A , H_c\right\} =\textrm{constant}
 \label{s2}
\end{equation}
where $\tau^{(0)}_{A}$ is the terminating element of any
constraint chain.

Returning to Eq. (\ref{d3}) for $n=1$, the generator
$G^{(1)}_{\alpha}$ is defined in Eq. (\ref{a20}). From Eq.
(\ref{s12}) the first and third terms in Eq. (\ref{a20}) vanish in
a simple way. According to Eq. (\ref{a22}), the remaining term
$\left\{\tau_\alpha ^{(0)}, H^{(1)}\right\}$ is proportional to a
summation of terms $ \left\{ \tau^{(0)}_\alpha , \tau^{(0)}_\beta
\right\}=\Delta_{\alpha\beta}$. Remembering that we have
considered systems with constant $\Delta$-matrix, we see that the
condition (\ref{s2}) results that $G_\alpha^{(1)}$ are constants
and $H^{(2)}$ is a function of $\eta$'s only. Hence, Eq.
(\ref{d3}) is also valid for $n=1$. Looking carefully at different
terms in Eq. (\ref{a21}) shows that under the considered
conditions the subsequent terms $G^{(n)}_{\alpha}$ for $n> 2$
vanish, giving finally
\begin{equation}
 H=H^{(0)}+H^{(1)}+H^{(2)}
 \label{s3}
\end{equation}
We see that the constancy of Poisson brackets of the second class
constraints and the Hamiltonian is sufficient to have an elegant
truncation of the embedded Hamiltonian. Moreover, it help's to
construct the embedding in such a way that the involuting algebra
of first class constraints with other constraints and with the
Hamiltonian is not violated. It should be noted that this
conclusion remains valid for BFT embedding with chain structure
\cite{shirmon2}, since it differs with abelian embedding only in
additional terms $\tau_{\alpha+1}^{(n)}$ in the definitions of
$G_\alpha^{(n)}$ which commute with first class constraints.

On the other hand, if in a certain model Eq. (\ref{s2}) does not
hold, then there is no guarantee that the embedding of second
class constraints is possible without violating the involuting
algebra of first class constraints. The problem is: what should we
do to satisfy (\ref{s2})? We will give our propositions to solve
this problem in the next section, specially for the Chern-Simons
theory. Before that let's take a look at this theory, its
constraint structure and the problem of its embedding.

The non-abelian Chern Simons theory in $(1+2)$ dimensions is
governed by the Lagrangian density \cite{chern}
\begin{equation}
{\cal{L}} =\frac{1}{2}k\varepsilon^{\mu\nu\rho}
(A^a_\mu\partial_\nu A^a_\rho + \frac{1}{3} f^{abc} A^a_\mu A^b
_\nu A^c_\rho)
 \label{a25}
 \end{equation}
where $A^a_\mu$ are dynamical fields, $f^{abc}$ are the structure
constants of some non-abelian Lie algebra,
$\varepsilon^{\mu\nu\rho}$ refer to the totally antisymmetric
tensor and $k$ is a constant. From the definition of canonical
momenta three (sets of) primary constraints emerge as follows
\begin{equation}
 \begin{array}{l}
 \Phi^{a 0}\equiv \pi^{a 0}\approx 0 \\ \Phi^{a i}\equiv \pi^{a i}-\frac{1}{2}k\varepsilon^{ij}A_j^a\approx
 0\hspace{2cm} i=1,2.
 \end{array}
 \label{a26}
 \end{equation}
The canonical Hamiltonian  can be written as
\begin{equation}
H_c=-k \int d^2{\mathbf{x}}
(A^a_0\varepsilon^{ij}\partial_iA^a_j+\frac{1}{6}
\varepsilon^{\mu\nu\rho}f^{abc}A^a_\mu A^b_\nu A^c_\rho).
 \label{a27}
 \end{equation}
The consistency condition of $\Phi^{a 0}$ gives the following
secondary constraint
\begin{equation}
\Phi^{a 3} \equiv k\varepsilon^{ij} \partial_i A^a_j +
\frac{k}{2}\varepsilon^{ij}f^{abc}A^b_iA^c_j\approx 0. \label{a28}
\end{equation}
No additional constraint is obtained from the consistency of the
constraints $\Phi^{a i}$ and $\Phi^{a 3}$. It seems that there
exist three second class constraints $\Phi^{a i}$ and $\Phi^{a
3}$, but one can combine the constraints to find two second class
and two first class constraints as follows
 \begin{equation}
\begin{array}{lll}
\Lambda^{a 0}=\Phi^{a 0} & \ \ \ \ \ \Lambda^{a 1}=\Phi^{a 1}
 & \ \ \ \ \ \Lambda^{a 2}=\Phi^{a 2}\\
\Lambda^{a 3}=\Phi^{a 3}+\partial_i\Phi^{a i} .
\end{array}
 \label{a29}
 \end{equation}
In Eqs. (\ref{a29}) $\Lambda^{a 1}$ and $\Lambda^{a 2}$ are second
class and $\Lambda^{a 0}$ and $\Lambda^{a 3}$ are first class
constraints. To find the redefinitions explained in Eqs.
(\ref{a29}) systematically we could first determine the unknown
Lagrange multipliers $\lambda^a_i$ in the total Hamiltonian
 \begin{equation}
 H_T=H_C+\lambda^a_0\Phi^{a 0}+\lambda_i^a \Phi^{a i}
 \label{a31}
 \end{equation}
and then use it for the consistency of the remaining constraint
$\Phi^{a0}$. In this way we find
 \begin{equation}
\lambda_i^a= \partial_i A^a_0+\frac{1}{2}f^{abc}A^b_i A^c_0.
 \label{a32}
 \end{equation}
Inserting $\lambda_i^a$ in the total Hamiltonian (\ref{a31}) gives
 \begin{equation}
H_T=H^{(0)} +  \lambda^a_0\Phi^{a 0},
 \label{a33}
 \end{equation}
 where
 \begin{equation}
H^{(0)}=H_C+\left( \partial_i A^a_0+\frac{1}{2}f^{abc}A^b_i
A^c_0\right) \Phi^{a i}.
 \label{s5}
 \end{equation}
Now the consistency of the primary constraint $\Phi^{a 0}$, using
this modified $H$,  gives
\begin{equation}
\Phi^{* a 3}=\left\{\Phi^{a 0} , H \right\}=\frac{k}{2}
\varepsilon^{ij}\partial_i A^a_j+\partial_i\pi^{a
i}+f^{abc}A^b_i\pi^{c i}
 \label{a34}
 \end{equation}
which is the same as $\Lambda^{a 3}$ in the definitions
(\ref{a29}). Since $\left\{ \Phi^{* a 3} , H \right\}=0$, no more
constraints would emerge. As is demonstrated in Eq. (\ref{a29}),
there are three constraint chains, one first class (including two
elements $\Lambda^{a0}$ and $\Lambda^{a3}$) and two second class,
each containing just one element. In fact, $\Lambda^{a1}$ and
$\Lambda^{a2}$ are the first and last elements of the
corresponding chains.

Suppose we want to construct an embedding for Chern-Simons theory.
The $\Delta$-matrix of second class constraints reads
\begin{equation}
\Delta^{ai,bj}({\mathbf{x},\mathbf{y}})\equiv
\left\{\Lambda^{ai}({\mathbf{x}},t) , \Lambda^{bj}({\mathbf{y}},t)
\right\}= \delta^{ab}\varepsilon^{ij}\delta
(\mathbf{x},\mathbf{y}) .
 \label{s6}
  \end{equation}
Since the $\Delta$-matrix is constant we can choose the finite
order embedding (\ref{s13}) as
\begin{equation}
 \tau^{a i}=\Lambda^{a i}+\eta^{a i} \hspace{1.5cm} i=1,2.
 \label{c10}
 \end{equation}
 The embedded Hamiltonian can also be found (see Eqs.
 \ref{a16}-\ref{a22}) as
 \begin{equation}
 H=H^{(0)}+H^{(1)} +H^{(2)}
 \label{c11}
 \end{equation}
 where
\begin{equation}
\begin{array}{l}
 H^{(1)}=\epsilon^{ij}\eta^{a}_{i} \partial_j A^a_0 +\frac{1}{2}f^{abc} \epsilon^{ij}\eta^{a}_{i} A^b_j A^c_0\\H^{(2)}=
 -\frac{1}{4}f^{abc}\eta^a_{i}\eta^{bi}A^{c}_{0}
\end{array}
 \label{c12}
 \end{equation}
As it is apparent the  constraints $\Lambda^{a0}$ and
$\Lambda^{a3}$ have no more vanishing Poisson brackets with the
embedded Hamiltonian even weakly. In other words, assuming
$\Lambda^{a0}=\pi^{a0}$ (for all $a$) as the primary constraints,
we will find some chains of second class constraints, due to
additional terms $H^{(1)}$ and $H^{(2)}$ in the Hamiltonian. This
shows that the initial gauge symmetry $A\rightarrow A+dA$
generated by the first class constraints $\Lambda^{a0}$ and
$\Lambda^{a3}$ is no more present in the embedded model.
Technically this has happened since $\Lambda^{a1}$ and
$\Lambda^{a2}$ as the last elements of the corresponding chains
have non-vanishing Poisson brackets with the Hamiltonian
(\ref{a33}). Therefore the requirement of vanishing the expression
given in Eq. (\ref{h2}) is not fulfilled. In fact, one may see
that $\left\{\Lambda^{ai} , H \right\}$ contain  terms with one or
two $A$-fields; hence, they do not commute with first class
constraints $\Lambda^{a0}$ and $\Lambda^{a3}$. Direct
investigation of the embedded Hamiltonian (\ref{c12}) also shows
that it no more commutes with the first class constraints of the
model. This is really the origin of the problem of BFT method for
some of the mixed constraint systems such as Chern-Simons.
\section{Solution}
In this section we give two different methods to overcome the
problem which lead to two different types of embedding for
Chern-Simons model.

$\mathbf{1)}$ In reference \cite{LoranShir1} some technics are
given which may help us satisfy the desired condition (\ref{s2}).
The main point is that, by adding terms which vanish on the
constraint surface one can redefine the constraints as well as the
Hamiltonian to satisfy Eq. (\ref{s2}). Suppose we are given two
second class chains terminating at non-commutating elements
$\Theta_{1}$ and $\Theta_2$ respectively, such that
\begin{equation}
 \begin{array}{l} \left\{ \Theta_{1} , \Theta_{2}
 \right\}=\delta \\
 \left\{\Theta_1 , H\right\}=\varrho  \\
 \left\{\Theta_2 , H\right\}=\gamma.
 \end{array}
 \label{h5}
 \end{equation}
Assume the following redefinitions
\begin{equation}
 \begin{array}{l}
 \hat{\Theta}_1=\gamma\Theta_1-\varrho\Theta_2 \\
 \hat{\Theta}_2= (\gamma)^{-1}\Theta_2 \ .
 \end{array}
 \label{h4}
 \end{equation}
 It is easy to observe that
\begin{equation}
 \begin{array}{l} \left\{ \hat{\Theta}_{1} , \hat{\Theta}_{2}
 \right\}\approx\delta \\
 \left\{\hat{\Theta}_1 , H\right\}\approx 0  \\
 \left\{\hat{\Theta}_2 , H\right\}\approx 1.
 \end{array}
 \label{s7}
 \end{equation}
In other words, the above redefinitions do not change the algebra
of second class constraints, while  their Poisson brackets with
Hamiltonian turn to be constants. The remainder of the problem is
straightforward. For non-abelian Chern-Simons theory this method
gives the following redefined constraints
 \begin{eqnarray}
\hat{\Lambda}^{a1}&=&
\left\{\frac{1}{2}k^2A^a_2(\partial_1A_0^a+f^{abc}A_0^cA_1^b)
-1\leftrightarrow2\right\} \\
&&-\left\{k\partial_1A_0^a\pi^{a1}+k\pi^{a1}f^{abc}A_0^cA_1^b
+1\leftrightarrow2\right\} \nonumber   \\
\hat{\Lambda}^{a2}&=& -(\pi^{a2}+\frac{1}{2}kA_1^a) /
(k\partial_1A_0^a+kf^{abc}A_0^cA_1^b),
 \label{s8}
 \end{eqnarray}
where no summation on the repeated index $a$ is assumed. Instead
of Eq. (\ref{c10}), the embedded constraints  are
 \begin{equation}
 \tau^{a i}= \hat{\Lambda}^{a i}+\eta^{a i}
 \hspace{1.5cm} i=1,2.
 \label{s9}
 \end{equation}
Finally according to Eq.(\ref{s3}), the embedded Hamiltonian is
 \begin{equation}
 H=H^{(0)}+\frac{1}{k} \sum_a \eta^{a1}
 \label{s10}
 \end{equation}
 where $H^{(0)}$ is given in Eqs. (\ref{s5}) and (\ref{a27}).

 $\mathbf{2)}$ By adding some auxiliary fields one can first convert the
first class constraints to second class ones and then the
traditional BFT method can be applied to the whole system. These
new auxiliary fields are different from those of the formal BFT
formalism. To see how this is possible, suppose we are given $K$
first class two-level chains originated from $K$ primary first
class constraints $\phi_i^{(0)} ; i=1,\cdots K$. We can assume
that $\phi_i^{(0)}$ are principally emerged, in some suitable
coordinates, from the definition of the momenta
\begin{equation}
p_i\equiv \frac{\partial L}{\partial\dot{q}_i}.
 \label{d6}
\end{equation}
One can easily see that the following extensions convert first
class constraints to second class ones:
\begin{equation}
\begin{array}{l}
p_i \rightarrow p_i+\xi_i \\
H_c \rightarrow H_c+\frac{1}{2}\sum_i p_{\xi_i}^2
\end{array}
 \label{d7}
\end{equation}
where $\xi_i$ and $p_{\xi_i}$ are auxiliary conjugate variables.
In the Lagrangian formalism this can be done by the replacement
\begin{equation}
L\rightarrow L-\xi_i \dot{q}_i+\frac{1}{2} \sum_i\dot{\xi_i}^2.
 \label{d8}
\end{equation}
It can be shown that the replacement (\ref{d8}) is in fact a gauge
fixing term inserted in the gauge invariant Lagrangian $L$. In
other words the new Lagrangian, or equivalently the new
Hamiltonian (\ref{d7}), gives the same equations of motion for
the gauge invariant quantities while destroys the arbitrariness
of the gauge dependent variables.

 Fortunately most physical models fall in the category of two level
 systems. However, for more
complicated systems it is not too difficult to add suitable
variables to convert first class constraints to second class ones.
For example, if there are four levels of constraints in a given
first class chain beginning with the momentum $p$, then by adding
two conjugate pairs $(\xi , p_\xi)$ and $(\eta , p_\eta)$ and the
replacements
\begin{equation}
\begin{array}{l}
p \rightarrow p+\xi \\
H_c \rightarrow H_c+\frac{1}{2} \eta^2 +p_\eta p_\xi ,
 \label{s11}
\end{array}
\end{equation}
one can convert the system to a second class one. In fact it is
not needed to give a detailed procedure for different cases which
may occur, since the process of constructing a second class system
from a first class one can be done easily for distinct models.

To apply this method to Chern-Simons theory one can make the
following replacement in the original Lagrangian (\ref{a25})
\begin{equation}
{\cal{L}}\rightarrow {\cal{L}}-\xi^a
\dot{A}^a_0+\frac{1}{2}\sum_{a}(\dot{\xi}^a)^2
 \label{c5}
\end{equation}
where $\xi^a$ are auxiliary fields in the configuration space. It
is obvious that the gauge symmetry $A\rightarrow A+df$ is lost in
the Lagrangian (\ref{c5}), while it can be shown that the gauge
invariant quantities are remained invariant. The Hamiltonian
(\ref{a27}) would consequently admit the following replacement
\begin{equation}
H_c\rightarrow H_c+\frac{1}{2}\sum_{a}(p_\xi^a)^2.
 \label{c6}
\end{equation}
By these replacements the primary and secondary constraints would
change to
\begin{equation}
\begin{array}{lll}
\hat{\Lambda}^{a0}\equiv \Lambda^{a0}+\xi^a \hspace{5mm}&
\hat{\Lambda}^{a1}\equiv \Lambda^{a1}
\hspace{5mm}&\hat{\Lambda}^{a2}\equiv \Lambda^{a2}\\
\hat{\Lambda}^{a3}\equiv\Lambda^{a3}+p_\xi^a.
\end{array}
 \label{c7}
\end{equation}
In this way we have a pure second class system for which the
ordinary finite order BFT method is applicable. The
$\Delta$-matrix now reads
 \begin{equation}
 \Delta^{a\mu,b\nu}({\mathbf{x},\mathbf{y})}= \left(
 \begin{array}{cccc}
 0&0&0&1 \\0&0&k&0 \\0&-k&0&0 \\ -1&0&0&0
 \end{array}\right) \delta^{ab}
 \delta ( \mathbf{x}-\mathbf{y})
 \label{c8}
 \end{equation}
 where $\mu ,\nu=0, \cdots 3$ are row and column
 indices of the above $4\times4$ matrix respectively. Again the
 $\Delta$-matrix is constant and one may write the following
finite extensions for the constraints
\begin{equation}
 \tau^{a \mu}=\hat{\Lambda}^{a \mu}+\eta^{a \mu} \hspace{1.5cm} \mu=0, \cdots 3.
 \label{s14}
 \end{equation}
    The embedded Hamiltonian can also be found (see Eqs.
 \ref{a16}-\ref{a22}) as
 \begin{equation}
 H=H^{(0)}+ \frac{1}{2} \sum _a (p_\xi^a)^2+H^{(1)} +H^{(2)}+H^{(3)}
 \label{c11}
 \end{equation}
 where $H^{(0)}$ is defined in Eq. (\ref{s5}) and
\begin{equation}
\begin{array}{l}
 H^{(1)}=\frac{-1}{k}\eta_1^a G_2^{a(0)}+\frac{1}{k}
 \eta_2^a G_1^{a(0)}+\eta_3^a G_0^{a(0)}\\H^{(2)}=
 \frac{-1}{k}\eta_1^a G_2^{a(1)}+\frac{1}{2k}
 \eta_2^a G_1^{a(1)}+\frac{1}{2}\eta_3^a G_0^{a(1)}
 \\ H^{(3)}=\frac{1}{3}f^{abc}(\eta_1^a \eta_2^b
 \eta_3^c +\eta_2^a\eta_3^b\eta_1^c)
\end{array}
 \label{c12}
 \end{equation}
in which
\begin{equation}
\begin{array}{l}
G_0^{a(0)}=k\epsilon^{ij}\partial_iA^a_j+\frac{k}{2}f^{abc}
\epsilon^{ij}A_i^bA_j^c\\G_i^{a(0)}=k\epsilon^{ij}
\partial_jA^a_0+k\epsilon^{ij}f^{abc}A^b_jA^c_0
\hspace{1.7cm} i=1,2  \vspace{0.3cm}\\
G_0^{a(1)}=\partial_i\eta_i^a+
f^{abc}(A_1^b\eta_1^c+A_2^b\eta_2^c)\\G_1^{a(1)}=f^{abc}
\eta_1^bA_0^c-k\partial_2\eta_3^a-kf^{abc}\eta_3^c
A_2^b\\G_2^{a(1)}=-f^{abc}\eta_2^bA_0^c+k\partial_1
\eta_3^a+kf^{abc}\eta_3^cA_1^b.
\end{array}
 \label{c13}
 \end{equation}
One can easily check that this Hamiltonian and the set of
constraints (\ref{s14}) construct a first class system.

It is worth noting that the above results are valid for abelian
Chern-Simons theory by imposing $f^{abc}=0$.

\section*{Concluding Remarks}
We showed that the BFT embedding method although applicable to
pure second class systems, is not guaranteed to work well for
systems possessing both first and second class constraints. The
Chern-Simons theory is a distinguished example in this regard. As
we saw, the bottle-neck condition is the requirement that at the
last level of consistency the second class constraints have
constant Poisson brackets with the Hamiltonian. This condition
guarantees that the algebra of first class constraints is not
violated during embedding of second class ones. However, we should
admit that this condition is actually stronger to some extent than
what is needed. In fact in concrete examples one may be able to
find different solutions in which the first class constraints
commute with the generators of the embedded Hamiltonian (i.e.
$G^{n}_{\alpha}$ in Eqs. \ref{a19}-\ref{a21}). So we think that
the problem is open in this regard.

However, if one insists that the critical condition (\ref{s2})
should be satisfied in any case, then several methods can be found
to redefine the constraints to reach this goal. We suggested just
one possibility in Eqs. (\ref{h4}). It may be possible to give
other (or better) solutions for this requirement. The problem is
also open in this direction. To sum up, in this approach one tries
to find the origin of this violation in the involuting algebra of
first class constraints and remove it.

As a second approach we gave another solution with a different
character. In this method we first convert the first class
constraints into second class ones by means of adding suitable
variables and then use the ordinary BFT method to embed the
resulting pure second class system into a first class one. It is
not usually a difficult task to construct a second class system
out of a first class one. We think that this will be easily done
in each concrete example and thus it is not needed to give general
prescriptions for that.

According to these methods, we gave two different types of
embedding for non-abelian Chern-Simons theory which includes the
abelian case easily by imposing $f^{abc}=0$. The embedding of the
abelian Chern-Simons theory was previously considered in
\cite{KimKim} by using an infinite number of auxiliary fields.
However, as far as we know, because of the mixed character of its
constraint structure, no finite order BFT embedding has been given
for Chern-Simons theory so far.
\section*{Acknowledgment}
The authors thank E. Mosaffa for reading the manuscripts.

\end{document}